\documentclass{elsarticle}

\usepackage[english]{babel}
\usepackage[utf8]{inputenc}
\usepackage{fancyhdr}

\usepackage[all]{xy}
\usepackage{eurosym}
\usepackage{tikz}
\usepackage{rotating}
\usepackage{amsmath}
\usepackage{amsfonts}
\usepackage{amssymb}
\usepackage{amsthm}
\usepackage{algorithm}
\usepackage{booktabs}

\usepackage{eurosym}
\usepackage{tikz}
\usepackage{rotating}

\usepackage{graphicx}
\usepackage{subfigure}
\usepackage{float}

\usepackage{parskip}

\usepackage{color}
\usepackage{ulem}

\definecolor{rojo}{rgb}{1,0,0}

\definecolor{rojooscuro}{rgb}{0.5,0,0}

\definecolor{verde}{rgb}{0,1,0}

\definecolor{verdeoscuro}{rgb}{0,0.5,0}

\definecolor{azul}{rgb}{0,0,1}

\definecolor{azuloscuro}{rgb}{0,0,0.5}

\newcommand{\bcas}{\begin{cases}}
\newcommand{\ecas}{\end{cases}}

\begin{document}

\begin{frontmatter}

\title{Financial option valuation by unsupervised learning with artificial neural networks}
\author{Beatriz Salvador$^{1}$, Cornelis W. Oosterlee$^{1,2}$, Remco van der Meer$^{1,2}$}
\address{$^{1}$ CWI – Centrum Wiskunde \& Informatica, Amsterdam, the Netherlands \\
         $^{2}$ DIAM, Delft University of Technology, Delft, the Netherlands}
\begin{abstract}
Artificial neural networks (ANNs) have recently also been applied to 
solve partial differential equations (PDEs). In this work, the classical problem of pricing European and American financial options, based on the corresponding PDE formulations, is studied. Instead of using numerical techniques based on finite element or difference methods, we address the problem using ANNs in the context of unsupervised learning. As a result, the ANN learns the option values for all possible underlying stock values at future time points, based on the minimization of a suitable loss function. For the European option, we solve the linear Black-Scholes equation, whereas for the American option, we solve the linear complementarity problem formulation. Two-asset exotic option values are also computed, since ANNs enable the accurate valuation of high-dimensional options. The resulting errors of the ANN approach are assessed by comparing to the analytic option values or to 
numerical reference solutions (for American options, computed by finite elements).
\end{abstract}


\end{frontmatter}

\section{Introduction}
The interest in machine learning techniques, due to the remarkable successes 
in different application areas, is growing exponentially.
Impressive results have been achieved in image recognition or natural language processing problems, among others. 
The availability of large data sets and powerful compute units has brought the broad field of data science to a next level.
ANNs are learning systems based on a collection of artificial neurons that constitute a connected network \cite{Zurada}. 
Such systems ``learn" to perform tasks, generally without being programmed with task-specific rules.
The neurons are organized in multiple layers; The input layer receives external data, the output layer produces the final result. The layers in between input and output are the so-called hidden layers \cite{book_ANN}.
Many different financial problems have also been addressed with machine learning, like stock price prediction, where ANNs are trained to detect patterns in historical data sets
to predict future trends \cite{Kryzanowski,Refenes}, or bond rating predictions, see \cite{Dutta,Moody,Singleton}. 

Motivated by the universal approximation theorems [3, 4], nowadays
ANNs are also being used to approximate solutions to  ordinary differential equations (ODEs) or partial differential equations (PDEs) [1,5-8].
We will contribute to this field by solving some PDEs that appear in computational finance applications with ANNs, following the unsupervised learning 
methodology introduced by \cite{Raissi} and refined in~\cite{Remco}. 
The resulting ANN-based methods do not require a discretization of the differential equation, and mesh generation is therefore not required.

The financial application on which we focus is the valuation of financial derivatives with PDEs. Generally, 
we can distinguish between  supervised and unsupervised machine learning techniques. Research so far has mainly focused on supervised machine learning, i.e. given input variables $x$ and labeled output variables $y$, the ANN is employed to learn the mapping function from the input to the output. The goal is then to approximate the mapping function accurately, so that for new input data $x'$, the corresponding output $y'$ is well approximated. 
Such ANN methodology usually consists of two phases. During the training phase, the ANN should learn the PDE solver with input parameters and output. This (off-line) phase usually takes substantial computing time. In the testing phase, the trained model is used to very rapidly approximate solutions for other parameter sets. 
In \cite{Grohs}, the authors showed that ANNs efficiently approximate the solution to the Black-Scholes equation. 
In \cite{Liu}, option values as well as the corresponding implied volatilities were directly computed with one neural network in a supervised learning approach. The authors in \cite{Amilon} examined whether an ANN could derive option pricing formulas based on market prices. ANN studies for American options are also found, like in \cite{Michael},
and in \cite{DGM}, where the option was formulated as a free boundary problem. In \cite{Alvaro} the American option implied volatility and implied dividend were assessed with
the help of ANNs.

The goal of the current work is to solve the financial PDEs by applying unsupervised machine learning techniques. 
In such a case, only the inputs of the network are known, and based on a suitable loss function that needs to be minimized, the ANN should ``converge'' to the solution of the PDE problem. 
The ANN should learn solutions that satisfy constraints that are imposed by the PDE and the boundary conditions, without using any information about the true solution.
These constraints are typically formulated as soft constraints, that are satisfied by minimizing some loss function. 
The potential advantage of applying ANNs to address PDE problems, instead of using classical numerical methods, is found in the problem's dimensionality. An ANN-based methodology does not suffer much from the curse of dimensionality. 
The authors of [1, 6, 7] provide evidence that for the well-known Poisson and Burgers equations, these unsupervised learning methods yield accurate results. 
The authors in \cite{Remco} extended the class of PDE solutions that may be approximated by these unsupervised learning methods, 
by translating the PDEs to a suitably weighted minimization problem for the ANNs to solve.
Moreover, in \cite{Becker, Becker2} American options were formulated as optimal stopping problems, where optimal stopping decisions were learned and so-called ANN regression was used to estimate the continuation values. 
This is an example of the unsupervised learning approach to solve a specific formulation of options with early-exercise features.

We will price European and American options modeled by the Black--Scholes PDE and look for solutions for all future time points and stock values. So, linear and nonlinear partial differential equations need to be solved. 
We will solve European and American option problems based on one and two underlying assets, as the methodology is easily extended to solving multi-asset options. 
For the European problems, the accuracy of the network can be measured as we have the analytic Black-Scholes solution as a reference. 
American options will be formulated as linear complementarity problems. Since an analytic solution is not known in this case, the reference solutions are obtained by finite element computations on fine meshes. 

This paper is organized as follows. In Section 2, the methodology to train the neural network is introduced. In Section 3, the financial PDE problems are formulated, for the linear and the nonlinear case. Numerical results, ANN convergence and solution accuracy, are presented in Section 4. Finally, Section 5 concludes.

\section{Artificial Neural Networks Solving PDEs}\label{methodologies}
In this section, we introduce the methodology following \cite{Remco} to solve linear and nonlinear time-dependent PDEs by ANNs. With this aim, we write a general PDE problem as follows:
\begin{align}\label{eqPDE}
\mathcal{N}_I(v(t,x)) &= 0, \quad x\in\widetilde{\Omega},\, t \in [0,T],\nonumber\\
\mathcal{N}_B(v(t,x)) &= 0 \quad \text{on}\,\, \partial\widetilde{\Omega},  \\
\mathcal{N}_0(v(t^{\ast},x)) &= 0 \quad x\in\widetilde{\Omega}\,\, \text{and}\,\, t^{\ast}=0 \,\,\text{or}\,\, t^{\ast}=T, \nonumber
\end{align}
where $v(t,x)$ denotes the solution of the PDE,  $\mathcal{N}_I(\cdot)$ is a linear or nonlinear time-dependent differential operator, $\mathcal{N}_B(\cdot)$ is a boundary operator, $\mathcal{N}_0(\cdot)$ is an initial or final time operator, $\widetilde{\Omega}$ is a subset of $\mathbb{R}^D$ and $\partial\widetilde{\Omega}$ denotes the boundary on the domain $\widetilde{\Omega}$.

As mentioned in the introduction, we will compute European and American option values for one and two underlying assets by unsupervised learning. 
The goal is to obtain $\hat{v}(t,x)$ by minimizing a suitable loss function $L(v)$ over the space of $k$-times differentiable functions, where $k$ depends on the order of the derivatives in the PDE, i.e
$$arg \min _{v\in \mathcal{C}^k}L(v) = \hat{v}\,,$$
where we  denote by $\hat{v}(t,x)$ the true solution of the PDE. 

Results are available that establish a relation between the value of the loss function and the accuracy of the approximated solution.
A general expression for the loss function, defined in terms of the $L^p$ norm, including a weighting, is defined as follows \cite{Raissi,Remco}:
\begin{align}\label{general_loss_function}
L(v) &= \lambda\int_{\Omega}\mid \mathcal{N}_I(v(t,x))\mid^p d\Omega \\
&+ (1-\lambda)\int_{\partial\Omega}\left(\mid \mathcal{N}_B(v(t,x))\mid^p +\mid \mathcal{N}_0(v(t,x))\mid^p \right) d\gamma,
\end{align}
where $\Omega = \widetilde{\Omega}\times [0,T]$, $\partial\Omega$ the boundary of $\Omega$  and 
\begin{align*}
&\mathcal{N}_I(v(t,x))\equiv N(v(t,x))-F(t,x) \quad\text{in}\,\,\Omega\,,\\
&\mathcal{N}_B(v(t,x))\equiv B(v(t,x))-G(t,x)  \quad\text{on}\,\,\partial\widetilde{\Omega}\,,\\
&\mathcal{N}_0(v(t^{\ast},x))\equiv H(x) - v(t^{\ast},x) \quad\text{in}\,\,\widetilde{\Omega}\times t^{\ast},\,\,\text{with} \,\, t^{\ast} = 0\,\,\text{or}\,\,  t^{\ast}= T.
\end{align*}
The integrals of the loss function are labeled as:
$$L_I(v) \equiv \int_{\Omega}\mid \mathcal{N}_I(v(t,x))\mid^p d\Omega,$$ and
$$L_B(v) \equiv \int_{\partial\Omega}\left(\mid \mathcal{N}_B(v(t,x))\mid^p + \mid \mathcal{N}_0(v(t,x))\mid^p\right) d\gamma,$$ 
which are denoted as the interior and the boundary loss functions, respectively.

Financial options with early-exercise features give rise to free boundary PDE problems.
Free boundary problems are well-known and often appearing in a variety of engineering problems. We recall some classical formulations of the free boundary problems that we encounter here:
\begin{itemize}
\item An optimal stopping time problem,
\item A linear complementarity problem (LCP),
\item A parabolic variational inequality,
\item A penalty problem.
\end{itemize}
We will focus on the reformulation of the free boundary problem as an LCP, and aim to solve this formulation by ANNs and unsupervised learning.
The generic LCP formulation reads,
\begin{align}\label{iqPDE}
\mathcal{N}_I(v(t,x))\cdot \mathcal{N}_0(v(t,x))  &= 0, \quad x\in\widetilde{\Omega}, t \in [0,T].
\end{align}
or, equivalently,
\begin{align*}
    \max(\mathcal{N}_{0}(v(t,x)), \mathcal{N}_{I}(v(t,x))) &= 0, \quad x\in\widetilde{\Omega},\, t \in [0,T],\nonumber\\
    \mathcal{N}_{B}(v(t,x)) &= 0, \quad \text{on}\,\, \partial\widetilde{\Omega},\\
    \mathcal{N}_{0}(v(t^\ast,x)) &= 0,\quad x\in\widetilde{\Omega}\,\, \text{and}\,\, t^{\ast}=0 \,\,\text{or}\,\, t^{\ast}=T.
\end{align*}


Our expression for the loss function, to solve the linear complementarity problem, is as follows: 
\begin{align}\label{loss_func_LCP}
    L(v) &= \lambda\int_{\Omega}\mid\max(\mathcal{N}_{0}(t,x,v), \mathcal{N}_{I}(t,x,v))\mid^pd\Omega \nonumber\\
& + (1-\lambda)\int_{\partial\Omega}\left(\mid \mathcal{N}_{B}(t,x,v)\mid^{p}+\mid \mathcal{N}_{0}(t,x,v)\mid^{p}\right)d\gamma\,.
\end{align}
As an alternative loss function for the LCP, a variance normalization loss function has also been considered~\cite{Remco}, which is defined as:
\begin{eqnarray}
L(v) &=& \frac{\int_\Omega\mid \max(\mathcal{N}_{0}(t,x,v),\mathcal{N}_{I}(t,x,v))\mid^p dx}{\int_\Omega (\max(\mid \mathcal{N}_{0}(t,x,v) \mid, \hat{\mathcal{N}}_{I}(t,x,v)))^p dx} \nonumber \\ &+& \frac{\int_{\partial \Omega}\left(\mid\mathcal{N}_{B}(t,x,v)\mid^p+\mid\mathcal{N}_{0}(t,x,v)\mid^p\right) d\gamma}{\int_{\partial\Omega}\mid v(t,x) - \bar{v}  \mid^p d\gamma},
\end{eqnarray}
where $\hat{\mathcal{N}}_{I}$ is defined as $\mathcal{N}_{I}$ but considering each term in absolute value and $\bar{v}$ is the mean of $v$ over the corresponding domain.

The parameter $\lambda\in (0,1)$ in the loss functions represents the relative importance of the interior and boundary functions in the minimization process. The choice of such value can be addressed in different ways, see \cite{Raissi,Remco}. 
In this work, the loss weight is, in most of the tests, set equal to $\lambda=0.5$. 
It was found in~\cite{Remco} that this choice works very well for PDE problems with smooth, non-oscillatory solutions (as we also encounter them in
the option valuation problems under consideration).
For some linear complementarity problems, we will compare the basic choice with the variance normalization loss function.
In addition, for some other cases, we will compute the loss function considering a so-called optimal loss weight (as in~\cite{Remco}).

Based on the loss function, the ANN has been trained with the Broyden-Fletcher-Goldfarb-Shanno optimization (BFGS). This is a quasi-Newton method which employs an approximate Hessian matrix. 
Particularly, we use the L-BFGS algorithm to optimize the vector $\theta$, which contains all parameters defining the neural network. 
The activation function used in the ANN is the hyperbolic tangent function $\tanh(x)$, however, other choices of the activation function can also be used, like the  sigmoid function (resulting in very similar results in this work). We will work with relatively small neural networks formed by four hidden layers with $20$ neurons each for the European and American options. Increasing the number of layers did not improve the accuracy of the solution significantly for these particular problems. Finally, the integral terms in the loss function are approximated by Monte Carlo techniques.

\section{Financial derivative pricing partial differential equations}\label{fin-pdes}
In this section the option pricing partial differential equation problems are presented. We briefly introduce the models.

\subsection{European options, one underlying asset}

The reference option pricing PDE for the valuation of a plain vanilla European, put or call, option is the Black-Scholes equation.
The underlying asset $S_t$ is assumed to pay a constant dividend yield $\delta$, and follows the geometric Brownian motion:
\begin{equation}\label{asset}
dS_t = (\mu-\delta) S_t dt + \sigma S_t dW^P_t\,,
\end{equation}
where $W^P_t$ is a Brownian motion. The drift term $\mu$, the risk-free interest rate, $r$, and the asset volatility, $\sigma$, are known functions. Assuming there are no arbitrage opportunities, the European option value follows from the Black--Scholes equation,
\begin{align}\label{B-S_eq_1d}
\begin{cases}
\mathcal{L}(v) = \partial_tv + \mathcal{A}v - rv = 0\,, \quad\quad S\in\widetilde{\Omega}\,\,, t\in[0,T)\,,\\
v(T,S) = H(S)\,,
\end{cases}
\end{align}
where operator $ \mathcal{A}$ is defined as,
\begin{equation}\label{operatorA}
 \mathcal{A}v \equiv \frac{1}{2}\sigma^2S^2\frac{\partial^2v}{\partial S^2} + (r-\delta)S\frac{\partial v}{\partial S}
\end{equation}
and function $H$ denotes the option's payoff, which is given by:
\begin{align}
\begin{cases}
(K-S)^+ \quad\quad \text{for a put option}\\
(S-K)^+ \quad\quad \text{for a call option}\,,
\end{cases}
\end{align}
with $K$ the strike price in the option contract.

In order to apply numerical methods to solve the PDE, a bounded domain should be considered and a proper set of boundary conditions should be imposed. We assume a domain large enough being $[0,S_{\infty}]$, with $S_{\infty}$ four times the strike $K$. Depending on the kind of option, call $v_c$ or put $v_p$, the problem (\ref{B-S_eq_1d}) is subject to the conditions:
\begin{align}\label{bc_1d}
 \begin{cases}
v_c(t,0) = 0\\
v_c(t,S_{\max}) = S_{\max} - Ke^{-r(T-t)}\,,
\end{cases}\quad\quad  \begin{cases}
v_p(t,0) = Ke^{-r(T-t)}\\
v_p(t,S_{\max}) = 0\,.
\end{cases}
\end{align}

The analytic solution for (\ref{B-S_eq_1d}) is known:
\begin{eqnarray*}
	v_c(t,S) &=& S\exp(-\delta (T-t))N_{0,1}(d_1) - K\exp(-r(T-t))N_{0,1}(d_2), \\
	v_p(t,S) &=& K\exp(-r(T-t))N_{0,1}(-d_2) - S\exp(-\delta(T-t))N_{0,1}(-d_1),
\end{eqnarray*}
with,
$$d_1 = \frac{\log(S/K) + (r-\delta + \sigma^2/2)(T-t)}{\sigma\sqrt{T-t}}\,\,,\,\,
d_2 = \frac{\log(S/K) + (r-\delta - \sigma^2/2)(T-t)}{\sigma\sqrt{T-t}}$$
and $N_{0,1}(x)$ the distribution function of a standard $\mathcal{N}(0,1)$ random variable.
Regarding the numerical solution with ANNs, we will use the methodology introduced in the previous section. In particular, the loss function is defined as: 
\begin{align}\label{loss_function}
L(v) 
&=\lambda \int_\Omega\mid\mathcal{L}(v(t,x))\mid^pd\Omega \nonumber\\
&+ (1-\lambda)\int_{\partial\Omega}\left(\mid v(t,x) - G(t,x)\mid^p + \mid v(t,x) - H(x)\mid^p\right) d\gamma,
\end{align}
where functions $G$ and $H$ denote the values of the spatial boundary conditions and final condition, respectively. The integral terms in the loss function are approximated by Monte Carlo techniques, as a result, we obtain the following interior and boundary loss function for the parameter vector $\theta$:
\begin{align}\label{variance normalization}
\widehat{L}(\theta) =& \lambda\frac{1}{n_I}\sum_{i=1}^{n_I}\mid \mathcal{L}(v(\mathbf{y}^{I}_{i},\theta)))\mid^p +\nonumber\\ &(1-\lambda)\left(\frac{1}{n_B}\sum_{i=1}^{n_B}\mid v(\mathbf{y}^{B}_{i},\theta) - G(\mathbf{y}^{B}_{i})\mid^p +\frac{1}{n_0}\sum_{i=1}^{n_0}\mid v(\mathbf{y}^{0}_{i},\theta) - H(\mathbf{x}^{0}_{i})\mid^p\right) .
\end{align} 
The collocation points $\{\mathbf{y}_{i}^{I}\}_{i= 1}^{n_{I}}$ and $\{\mathbf{y}_{i}^{B}\}_{i=1}^{n_{B}}$ are uniformly distributed over the domain $\Omega$ and the boundary $\partial\widetilde{\Omega}$ and $\{\mathbf{y}_{i}^{0}\}_{i=1}^{n_{0}}$ are uniformly distributed over the domain $T\times \widetilde{\Omega}$  , respectively and $\mathbf{y} = (t,x)$.

\subsection{Two underlying assets}\label{sub_2asset_europ}
We extend the model for one underlying asset to valuing basket options with two underlying assets. 
The two-asset prices follow the following dynamics,
\begin{align*}
dS_{1_t} &= (\mu_1-\delta_1)S_{1_t} dt + \sigma_1 S_{1_t} dW_t^1,\\
dS_{2_t} &= (\mu_2-\delta_2)S_{2_t} dt + \sigma_2 S_{2_t} dW_t^2,
\end{align*}
where $\mu_1,\mu_2$ are drift terms, $\delta_1,\delta_2$ dividend yields, the Brownian increments, $dW^i$ for $i = 1,2$, satisfy $\mathbb{E}(dW^i) = 0$, and
the underlying assets are correlated:
$$\text{corr}(W^1,W^2) = \rho t \quad \quad \text{or} \quad \quad\mathbb{E}(dW^1,dW^2) =\rho dt\,.$$

In the Black-Scholes framework, the two-asset European option price, $v(t,S_1,S_2)$, satisfies the following PDE:
\begin{align}\label{2asset_European}
\begin{cases}
\mathcal{L}_2(v) = \partial_t v + \mathcal{B}v - rv = 0\quad (S_1,S_2)\in \widetilde{\Omega}\,,\quad t\in [0,T),\\
v(T,S_1,S_2) = H_2(S_1,S_2)\,,
\end{cases}
\end{align}
where the operator $\mathcal{B}$ is defined as follows:
\begin{eqnarray}\label{operatorB}
 \mathcal{B}v &\equiv& \frac{1}{2}\sigma_1^2S_1^2\frac{\partial^2v}{\partial S_1^2} + \frac{1}{2}\sigma_2^2S_2^2\frac{\partial^2v}{\partial S_2^2}  + \rho\sigma_1\sigma_2 S_1 S_2\frac{\partial^2v}{\partial S_1 \partial S_2}\nonumber \\
&+& (r-\delta_1)S_1\frac{\partial v}{\partial S_1} + (r-\delta_2)S_2\frac{\partial v}{\partial S_2},
\end{eqnarray}
 and function $H_2(S_1, S_2)$ denotes the payoff function. By prescribing different payoff functions, different options can be defined, like an
exchange option, rainbow option or an average put option. We will deal with the exchange option, for which an analytic solution is given by the Margrabe's formula \cite{Margrabe} and the max-on-call rainbow option, for which a closed-form expression was introduced in \cite{Johnson} and \cite{Stulz}. These particular options are defined by their payoff functions:
\begin{eqnarray*}
	H_2(S_1,S_2) &=& (S_1 - S_2)^+\, \mbox{ exchange option}, \\
	H_2(S_1,S_2) &=& (\max(S_1,S_2)-K)^+\, \mbox{ max-on-call rainbow option}.
\end{eqnarray*} 

According to the Margrabe's formula, the fair value of a European exchange option at time $t$ is given by:
\begin{equation}\label{Margrabe_formula}
v(t,S_1,S_2) = e^{-\delta_1(T-t)}S_1(t)N_{0,1}(d_1) - e^{-\delta_2(T-t)}S_2(t)N_{01}(d_2)
\end{equation}
where $N_{0,1}$ again denotes the cumulative distribution function for the standard normal, $\sigma = \sqrt{\sigma_1^2 + \sigma_2^2-2\sigma_1\sigma_2\rho}$ and
$$d_1 = (\log(S_1(t)/S_2(t)) + (\delta_2 - \delta_1 + \sigma^2/2)T/\sigma\sqrt{T-t} \quad,\quad d_2 = d_1-\sigma\sqrt{T-t}\,.$$

With the following parameters:
$$d_i = \frac{\log(S_i/k)+(r-\delta_i+\frac{\sigma_i^2}{2})(T-t)}{\sigma_i\sqrt{T-t}}\,,$$
$$\rho_1 = \frac{\sigma_1 - \rho\sigma_2}{\sigma}\quad\text{and}\quad \rho_2 = \frac{\sigma_2 - \rho\sigma_1}{\sigma}\,,\quad i = 1,2,$$ the closed-form formula for a call on the maximum is given by:
\begin{align}\label{max_call_formula}
v_c^{\max}(t,S_1,S_2) &= S_1e^{-\delta_1(T-t)}M(d_1,d;\rho_1) + S_2e^{-\delta_2(T-t)}M(d_2,-d+\sigma\sqrt{T-t};\rho_2)\nonumber\\
&-Ke^{-r(T-t)}(1-M(-d_1+\sigma_1\sqrt{T-t},-d_2+\sigma_2\sqrt{T-t};\rho)),
\end{align}
where $M$ is the cumulative bivariate normal distribution
$$M(a,b;\rho) =\frac{1}{2\pi\sqrt{1-\rho^2}}\int_{-\infty}^{a}\int_{-\infty}^{b}e^{-\frac{x^2-2\rho xy + y^2}{2(1-\rho)}}dx dy\,.$$

To obtain a numerical solution of the PDE (\ref{2asset_European}), we bound the domain and impose appropriate boundary conditions. The computational domain should be sufficiently large, $[0,S_{1\infty}]\times[0,S_{2\infty}]$, where $S_{1\infty}=S_{2\infty}=4K$ ($K$ the option strike). In the particular case of the exchange and rainbow max-on-call options, where the analytic solutions are known, we impose as boundary conditions the analytic option value on each boundary. 

Similar to the one-dimensional problem, we address the European exchange option problem building the loss function as a sum of the interior and boundary loss functions, using $\lambda=0.5$.
\subsection{American options, one underlying asset}
As we have introduced in Section \ref{methodologies}, we also address the problem for an American option depending on one underlying asset price. With this aim, we focus on the linear complementarity formulation.
%
%
%
%
%
\subsubsection{Linear complementarity formulation}
We will here consider the linear complementarity problem (LCP) American option valuation formulation, see, for example, \cite{Wilmott, Ikonen2}, as follows,
\begin{align}
\begin{cases}
\mathcal{L}(v) = \partial_tv + \mathcal{A}v - rv \leq 0\,, \quad\quad S\in\widetilde{\Omega}\,, t\in[0,T)\,,\\
v(t,S) \geq H(S),\\
\mathcal{L}(v)(v-H) = 0,\\
v (T,S) = H(S)\,.
\end{cases}
\end{align}
This LCP can be rewritten as a nonlinear PDE as follows
\begin{align}\label{LCP_formulation}
\begin{cases}
\max\{H(S) - v(t,S),\mathcal{L}(v)\}=0 \,,\quad\quad S\in\widetilde{\Omega}\,, t\in[0,T)\,,\\
v(T,S) = H(S)\,.
\end{cases}
\end{align}
%
%
%
%
Essentially, using the same methodology for solving the European option PDEs, we address the linear complementarity formulation and its equivalent formulation as a nonlinear PDE given by (\ref{LCP_formulation}). 

As we introduced in Section \ref{methodologies}, the loss function can be formulated using variance normalization. Moreover, in case of the American option we will also compute $\lambda$ as the optimal loss weight.  

The loss function based on variance normalization depends on the variance of the network output. For the Black-Scholes American option problem, the loss function following variance normalization is given by
\begin{eqnarray}
L(v) &=& \frac{\int_\Omega\mid \max(H(x) - v(t,x),\mathcal{L}(v(t,x)))\mid^p dx}{\int_\Omega (\max(\mid H(x) - v(t,x) \mid, \tilde{\mathcal{L}}(v(t,x))))^p dx} \nonumber \\
&+& \frac{\int_{\partial \Omega}\left(\mid v(t,x) - G(t,x)\mid^p + \mid v(t,x) - H(x)\mid^p\right) d\gamma}{\int_{\partial\Omega}\mid v(t,x) - \bar{v}  \mid^p d\gamma},
\end{eqnarray}
 where $\tilde{\mathcal{L}}(v(t,x)))$ is defined as follows
 \begin{equation}\label{Lv_abs}
\mathcal{\tilde{L}}(\hat{v}) = \mid \partial_t \hat{v} \mid + \mid \frac{1}{2}\sigma^2S^2\partial^2_{SS}\hat{v}\mid + \mid (r-\delta)S\partial_S \hat{v}\mid + \mid r\hat{v} \mid \,,
\end{equation}
function $G$ refers to the boundary conditions imposed in a bounded domain which are defined as in (\ref{bc_1d}) and function $H$ denotes the final condition. Moreover, $\bar{v}$ is the mean of $v$ over the corresponding domain, which is given as 
 \begin{equation}
 \bar{v} = \frac{1}{\| \partial \Omega \|}\int_{\partial \Omega}v(t,x) d\Omega\,.
 \end{equation}
Then, approximating each integral term by Monte Carlo techniques the resulting function is defined as follows
\begin{align}\label{variance normalization_2}
\widehat{L}(\theta) =& \frac{\sum_{i=1}^{n_I}\mid\max(H(x^{I}_{i})-v(\mathbf{y}^{I}_{i},\theta), \mathcal{L}(v(\mathbf{y}^{I}_{i},\theta)))\mid^p}{\sum_{i=1}^{n_I}\max(\mid H(x^{I}_{i})-v(\mathbf{y}^{I}_{i},\theta)\mid, \mathcal{\tilde{L}}(v(\mathbf{y}^{I}_{i},\theta)))^p} +\nonumber\\ &\frac{\frac{1}{n_B}\sum_{i=1}^{n_B}\mid v(\mathbf{y}^{B}_{i},\theta) - G(\mathbf{y}^{B}_{i})\mid^p +\frac{1}{n_0}\sum_{i=1}^{n_0}\mid v(\mathbf{y}^{0}_{i},\theta) - H(\mathbf{x}^{0}_{i})\mid^p  }{\frac{1}{n_\ast}\sum_{i=1}^{n_\ast}\mid v(\mathbf{y}^{\ast}_{i},\theta) - \frac{1}{n_\ast}\sum_{j=1}^{n_\ast}v(\mathbf{y}^{\ast}_{j})\mid^p},
\end{align} 
with $\theta$ containing all parameters of the neural network, vector $\mathbf{y} = (t,x)$, and the collocation points $\{\mathbf{y}_{i}^{\ast}\}_{i= 1}^{n_{\ast}}$ are uniformly distributed over the boundary $\partial\Omega$.

An alternative is to build the loss function based on an optimal loss weight. However, optimizing $\lambda$ can be nontrivial.

In order to find the optimal loss weight, we may look for a so-called $\epsilon$-close solution to the true solution $\hat{v}$, see~\cite{Remco},
$$ \biggl| \frac{\partial^n v}{\partial y_i^n} - \frac{\partial ^n \hat{v}}{\partial y_i^n}\biggr|\leq \epsilon \frac{\partial ^n \hat{v}}{\partial y_i^n},$$
for all $n\geq 0$ and $i\in {1,\ldots,d}$, where $d$ is the dimension of the problem. 
Satisfying such condition, the value of the optimal loss weight $\lambda^*$ should be:
\begin{equation}\label{opt_loss_weight}
\lambda^* = \frac{\int_{\partial\Omega}\mid\hat{v}(t,x)\mid^p d\gamma}{\int_{\Omega} ( \hat{\mathcal{N}}_I(t,x,\hat{v}))^p d\Omega + \int_{\partial\Omega}\mid\hat{v}(t,x)\mid^p d\gamma },
\end{equation}
where function $\hat{\mathcal{N}}_I(x,\hat{v})$ is defined as the function $\mathcal{N}_I(x,\hat{v})$ with each term in absolute value.
This expression of $\lambda^*$ is constant when the analytical solution is known. However, for the American options where the analytical solution is not known, the optimal loss weight can be computed by approximating the value of $\hat{v}$  in (\ref{opt_loss_weight}) by the trained solution. Note that in this case, the loss weight is a function instead of a constant value and is optimized by the neural network.


As a result, the loss function is built in the following way:
\begin{align}
L(v) =& \lambda^* L_I(v) + (1- \lambda^*)L_B(v)\nonumber\\
=&\lambda^* \int_\Omega\mid\max(H(x)-v(t,x),\mathcal{L}(v(t,x)))\mid^pd\Omega +\nonumber\\
&(1-\lambda^*)\int_{\partial\Omega}\left(\mid v(t,x) - G(t,x)\mid^p + \mid v(t,x) - H(x)\mid^p\right)d\gamma,
\end{align}
and the optimal loss weight is given in terms of the trained solution  $v$, as follows:
$$\lambda^* = \frac{\int_{\partial\Omega}\mid v(t,x)\mid^pd\gamma}{\int_\Omega (\mathcal{\tilde{L}}_1(v))^p d\Omega + \int_{\partial\Omega}\mid v(t,x)\mid^pd\gamma},$$
with $$\mathcal{\tilde{L}}_1(v) = \max(\mid H(x) - v(t,x)\mid, \mathcal{\tilde{L}}(v)),$$
where $\mathcal{\tilde{L}}(v)$ is defined as in (\ref{Lv_abs}).

\subsection{Two-Asset American option}

The one underlying asset American option pricing problem is extended to also price multi-asset American options. We focus on two underlying assets and formulate the problem as a linear complementarity problem.  Based on two asset prices  following correlated geometric Brownian motion, the American option value can be modeled by the following linear complementarity problem:
\begin{align}\label{2asset_American}
\begin{cases}
\mathcal{L}_2(v) = \partial_t v + \mathcal{B}v - rv \leq 0 \,, \quad\quad (S_1,S_2)\in\widetilde{\Omega}\,, t\in [0,T),\\
v(t,S_1,S_2) \geq H_2(S_1,S_2),\\
\mathcal{L}_2(v)(v-H_2)=0,\\
v(T,S_1,S_2) = H_2(S_1,S_2)\,.
\end{cases}
\end{align}
Operator $\mathcal{B}$ is defined as in (\ref{operatorB})  and function $H_2(S_1, S_2)$ denotes the payoff function. In order to compare with the European option problem, an American call on the maximum is also priced, moreover, we address a two-asset spread option and a put arithmetic average option. Then, the payoff functions are defined as: 
\begin{eqnarray*}
	H_2(S_1,S_2) &=&  (\max(S_1,S_2)-K)^+,\;\, \mbox{max-on-call rainbow option}, \\
	H_2(S_1,S_2) &=&(S_1 - S_2 - K)^+,\;\, \mbox{ asset spread option}, \\
	H_2(S_1,S_2) &=& (K - (S_1 + S_2)/2)^+,\;\, \mbox{arithmetic average put}.\\
\end{eqnarray*} 
In order to solve the linear complementarity formulation using numerical methods, a bounded domain should be considered and appropriate boundary conditions should be imposed. In particular, we consider a domain large enough to avoid that the solution is affected by the conditions, in the interested regions of the asset prices.
Whereas for the European option problem the analytical solution is known and imposed as a boundary condition, for the American options problem, where the analytical solution is not known, we should define the appropriate boundary conditions. Then, we start studying at which boundaries a condition should be imposed. Following \cite{Oleinik}, that includes the theory of Fichera \cite{fic1960}, we introduce the notation $x_0=\tau$, $x_1=S_1$, $x_2=S_2$, and the domain $\Omega^\ast = (0,x_0^{\infty})\times(0,x_1^{\infty})\times (0,x_{2}^{\infty})$, where $x_{0}^{\infty} = T$, $x_{1}^{\infty} = S_{1\infty}$ and $x_{2}^{\infty} = S_{2\infty}$.
The boundary of $\Omega^{\ast}$ is,
$$\partial\Omega^{\ast} = \bigcup_{i=0}^{2}(\Gamma_{i}^{\ast,-}\cup \Gamma_{i}^{\ast,+}),$$
where we use the notation:
$$\Gamma_{i}^{\ast,-} = \{(x_{0},x_{1},x_{2})\in\partial\Omega^{\ast}\,, \/ x_{i}=0\}, $$
$$\Gamma_{i}^{\ast,+} = \{(x_{0},x_{1},x_{2})\in\partial\Omega^{\ast}\,, \/ x_{i}=x_i^\infty\}\,.$$

Then, the PDE in (\ref{2asset_American}) can be written in the form:
$$\sum_{i,j=0}^{2}b_{i,j}\frac{\partial^2 v}{\partial x_{i}\partial x_{j}} + \sum_{j=0}^{2}p_{j}\frac{\partial v}{\partial x_{j}} + c_0 v \leq g_0,$$
where the involved data are defined as follows:

$$B(x_0,x_1,x_2) = (b_{ij}) = \left(\begin{matrix}
0 & 0 & 0\\
0 & \frac{1}{2}\sigma_1^2x_1^2 & \frac{\rho\sigma_1\sigma_2 x_1 x_2}{2}\\
0 & \frac{\rho\sigma_1\sigma_2 x_1 x_2}{2} & \frac{1}{2}\sigma_2^2 x_2^2
\end{matrix}\right)\,,\quad\quad c_0 = r\,,$$
$$p(x_0,x_1,x_2) = (p_j) = \left(\begin{array}{c}
-1\\
(r-\delta_1)x_1\\
(r-\delta_2)x_2\end{array}\right)\,,\quad\quad g(x_0,x_1,x_2)=0\,.$$
Next, we introduce the following subset of $\Gamma^{\ast}$ in terms of the normal vector to the boundary pointing inwards $\Omega^{\ast}$, $\overrightarrow{m} = (m_0,m_1,m_2)$
$$\Sigma ^{0} = \left\{x\in\partial\Omega^{\ast} / \sum_{i,j=0}^{2} b_{ij}m_im_j = 0\right\}\quad,\quad \Sigma ^{1} = \partial\Omega^{0} - \Sigma^{0}\,,$$
$$\Sigma^{2} = \left\{x\in \Sigma^{0} / \sum_{i=0}^{2} \left(p_i - \sum_{j=0}^{2}\frac{\partial b_{ij}}{\partial x_j}\right)m_i\leq 0\right\}\,.$$

In this particular problem, we have: $\Sigma^0 = \Gamma_0^{\ast,-}\cup \Gamma_0^{\ast,+} \cup\Gamma_1^{\ast,-}\cup \Gamma_2^{\ast,-}$, $\Sigma^{1}  = \Gamma_1^{\ast,+}\cup \Gamma_2^{\ast,+}$ and $\Sigma^2 = \Gamma_0^{\ast,+}$. Thus, following \cite{Oleinik}, the boundary conditions must be imposed over the subset $\Sigma^{1}\cup \Sigma^{2}$ which matches with the set $\Gamma_0^{\ast,-}\cup \Gamma_1^{\ast,+}\cup \Gamma_2^{\ast,+}$. Then, is not necessary to impose boundary conditions above the boundary where the asset prices $S_1$ and $S_2$ are equal zero. Moreover, for simplicity, we assume that the option value is equal to the payoff when the asset prices $S_1$ and $S_2$ take the maximum values.

Next, taking into account the methodologies proposed to solve the one-dimensional American problem and the two-dimensional European problem, we propose the loss function to solve the multi-asset American option by artificial neural network. First of all, we rewrite the linear complementarity problem (\ref{2asset_American}) as the equivalent nonlinear PDE: 
\begin{align}\label{2asset_LCP_formulation}
\begin{cases}
\max\{H_2(S_1,S_2) - v(t,S_1, S_2),\mathcal{L}_2(v)\}=0,\\
v(T,S_1, S_2) = H_2(S_1, S_2)\,.
\end{cases}
\end{align} 

Similar to the previous problems, we build the loss function as the sum of the interior and boundary loss functions as follows:
\begin{align}\label{american_loss_2d}
L(v) =& \lambda L_I(v) + (1- \lambda)L_B(v)\nonumber\\
=&\lambda \int_\Omega\mid\max(H_2(\mathbf{x})-v,\mathcal{L}(v))\mid^pd\Omega +\nonumber \\
&(1-\lambda)\int_{\partial\Omega}\left(\mid v(t,\mathbf{x}) - G(t,\mathbf{x})\mid^p + \mid v(t,\mathbf{x}) - H(\mathbf{x})\mid^p\right) d\gamma,
\end{align}
where function $G$ refers to the boundary conditions and function $H$ denotes the final condition imposed for the problems. Note that the loss function is a generalization of the loss function introduced for the one asset problem and the integral terms are also approximated by Monte Carlo techniques.

\section{ANN Option Pricing Results}
In this section the European and American options values are computed with the ANNs based on the loss functions introduced. We apply the unsupervised learning methodology from
the previous section to compute the solutions and show some results. For the following tests, we have considered the parameter $p = 2$ in the loss functions.

\subsection{European options}
First of all, we discuss the European option single asset results obtained solving the PDE problem (\ref{B-S_eq_1d}) by ANNs. 

The results are presented with the loss function introduced in (\ref{loss_function}) and here we use the basic choice $\lambda = 0.5$. 
Recall that optimal loss weight-based loss function  or the variance normalization technique are especially useful in the case of nontrivial solutions.

\begin{figure}[ht!]
  \centering 
      \includegraphics[width=0.45\linewidth]{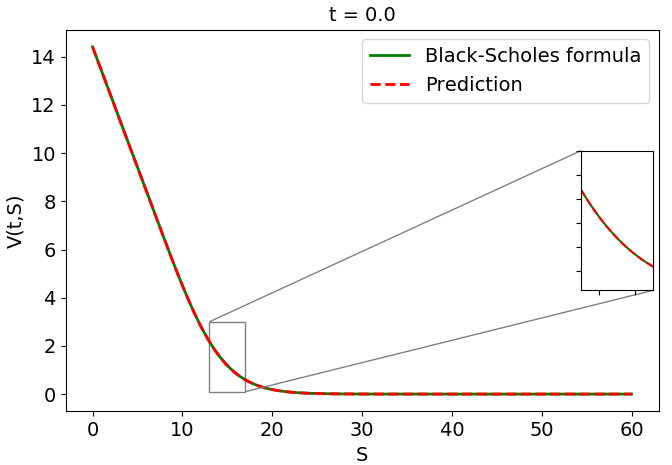}  
    \includegraphics[width=0.45\linewidth]{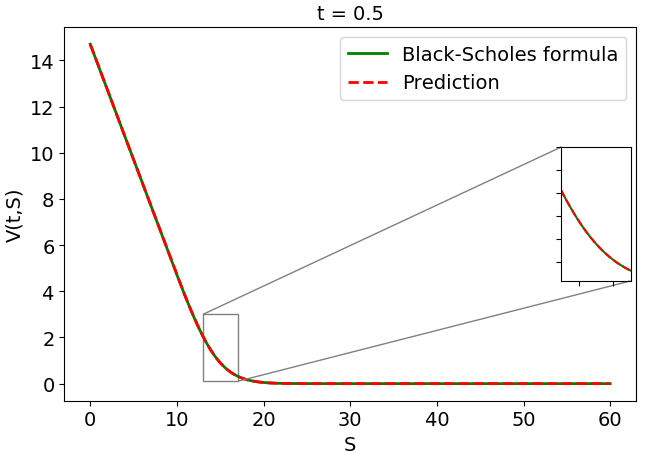}
    \caption{European put option for different times instances, $t=0, t=0.5$, with $\lambda = 0.5$.}
      \label{fig:ANN_vs_FEM_l05}
\end{figure}

We start with a European put option, with the following parameters values: $\sigma = 0.25$, $r = 0.04$, $T = 1$, $K = 15$, $S_\infty = 4K$, $\delta = 0.0$. In Figure \ref{fig:ANN_vs_FEM_l05}, the ANN-based, trained and the analytical solution are plotted for two time instances.

We measure the accuracy of the solution generated by the ANN by comparing the relative error of the trained solution $v_{ANN}$  with the analytic solution $v_{BS}$, as follows:
\begin{equation}\label{error}
error = \frac{\Vert  v_{BS}-v_{ANN}\Vert_{L^2}}{\Vert v_{BS} \Vert_{L^2}}\,.
\end{equation}
\begin{figure}[ht!]
  \centering 
    \includegraphics[width=0.45\linewidth]{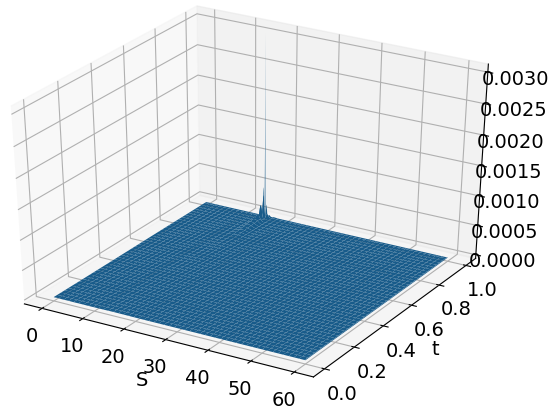}
    \caption{Error surface for the ANN solution.}
      \label{fig:Error_surface}
\end{figure}
In Figure \ref{fig:Error_surface} the error throughout the domain is plotted. Clearly, the biggest error in the ANN solution is found close to the strike price at maturity time $t=T$, where the payoff is non-smooth.
The relative error according to (\ref{error}) with $\lambda = 0.5$ is equal to $2.23\times 10^{-4}$.

Next, we show some results for a European option depending on two underlying assets.

\begin{figure}[ht!]
  \centering 
    \includegraphics[width=0.45\linewidth]{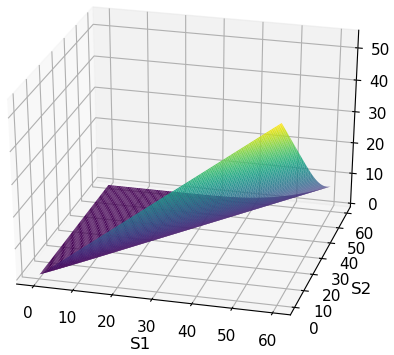}  
    \includegraphics[width=0.45\linewidth]{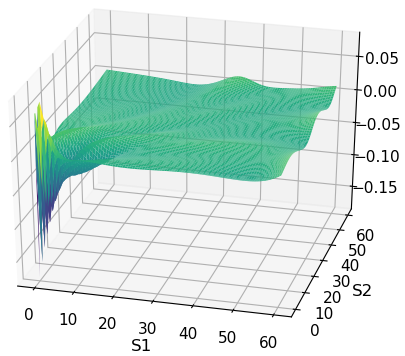}
    \caption{European exchange option, with parameters: $\sigma_1 = \sigma_2 = 0.25$, $\rho = 0.1$, $r = 0.05$, $\delta_i= 0.1$, $S_{1\infty} = S_{2\infty} = 60$ and loss weight $\lambda = 0.5$. }
      \label{fig:Europ_asset_exchange}
\end{figure}

The corresponding loss function has been optimized by means of the L-BFGS algorithm and choosing the $\tanh$ as the activation function. 
In the last layer a linear activation function is considered. 
We have plotted in Figure \ref{fig:Europ_asset_exchange} the ANN solution for the European exchange option. The error, comparing the approximated ANN solution with the analytical solution given by (\ref{Margrabe_formula}) is also plotted. Note that the maximum error is obtained for the minimum value of both asset prices, which is related to $S_1/S_2$ in the expression of $d_1$ in (\ref{Margrabe_formula}).

Due to the relatively big differences in the asset prices $S_1$, $S_2$ and the time $t$-values, we have scaled the inputs of the artificial neural network, i.e. the original domain $\widetilde{\Omega} = [0,S_{1\infty}]\times[0,S_{2\infty}]$, is scaled to a dimensionless computational domain, i.e.,  $\widetilde{\Omega}^{\ast} = [0,1]\times[0,1]$. 
By pricing the option with the parameters, $S_{1\infty} = S_{2\infty} = 4K$, $\sigma_1 = \sigma_2 = 0.25$, $\rho = 0.1$, $r_R = 0.04$, $r = 0.3$ and $T= 1$,  for several values of $K$, modifying the original domain, we found that scaling the input parameters is not sufficient to obtain highly accurate results for large domain sizes. 
In Table \ref{accurate_domain}, the error for a European max-call option is presented, based on different unscaled domain sizes. It can be observed that as the domain increases the accuracy of the neural network solution decreases. 


\begin{table}
\begin{center}
 \begin{tabular}{||c | c  ||} 
 \hline
 ($S_{1,\infty} , S_{2,\infty}$) & Relative error   \\ [0.5ex] 
 \hline\hline
 (10 , 10) & $2.58\times 10^{-4}$   \\ 
 \hline
 (60 , 60) & $3.17\times 10^{-4}$   \\ 
 \hline
 (120 , 120) & $8.08\times 10^{-4}$  \\ 
 \hline
 (180 , 180) & $1.71\times 10^{-2}$  \\ 
 \hline
 (240 , 240) & $2.75\times 10^{-1}$ \\ 
 \hline
 (300 , 300) & $3.96\times 10^{-1}$  \\ 
 \hline
 (360 , 360) & $4.30\times 10^{-1}$  \\ 
 \hline

\end{tabular}
\caption{Relative error for different domains.}
\label{accurate_domain}
\end{center}
\end{table}

In order to understand, the reasons for the degraded accuracy with an increasing  domain size, we have computed the gradients of the interior and boundary loss functions. In Table \ref{gradient_table}, we present these values for the European max-call. The gradient of the interior loss remains constant, note that the domain is always $[0,1]\times[0,1]$, however, the gradient of the boundary loss increases with the size domain. Clearly, the interior and boundary loss functions do not have the same dependency on the domain size. 
\begin{table}
\begin{center}
 \begin{tabular}{|| c | c | c  ||} 
 \hline
 ($S_{1,\infty} , S_{2,\infty}$)& $\|\partial L_I/\partial \omega\|_{L^2}$& $\|\partial L_B/\partial \omega\| _{L^2} $ \\ [0.5ex] 
 \hline\hline
 (10 , 10) &$0.4325$ & $8.8515$   \\ 
 \hline
 (60 , 60) & $0.4325$ &$52.6274$   \\ 
 \hline
 (120 , 120) & $0.4325$ & $105.1598$  \\ 
 \hline
 (180 , 180) & $0.4325$  & $157.6923$  \\ 
 \hline
 (240 , 240) & $0.4325$ & $210.2249$ \\ 
 \hline
 (300 , 300) & $0.4325$  & $262.7574$  \\ 
 \hline
 (360 , 360) & $0.4325$ & $315.2899$  \\ 
 \hline

\end{tabular}
\caption{Gradient values for different domain sizes with standard weights.}
\label{gradient_table}
\end{center}
\end{table}

We wish to compute accurate approximations of the solution independent the domain size, and therefore the ANN needs to be modified. The initialization of the weights is adapted by using a variation of the Xavier initialization. In particular, the initial values of the weight values in the last layer of the ANN will be scaled, by multiplying them by the maximum option value. 
As a result, we obtain a solution which is accurate independent of the size of the domain, see Table \ref{accurate_domain_scale_weights}.
This adaptation, i.e. the weights having similar magnitude as the expected largest option value in the output, forms a robust weight initialization. Moreover, such initialization helps for the interior and boundary loss functions to have similar sensitivity to the domain size. In Table \ref{gradient_table_new}, we can observe such behaviour, where the rate between both gradients remains close to $1/3$ when the size of the domain increases. 
Our results show that the BFGS optimization doesn't seem to pick up the gradient if the initial weights are not sufficiently large. Moreover, similar results can be observed when the inputs are not scaled.
\begin{table}
\begin{center}
 \begin{tabular}{||c | c  ||} 
 \hline
 ($S_{1,\infty} , S_{2,\infty}$) & Relative error   \\ [0.5ex] 
 \hline\hline
 (10 , 10) & $3.60\times 10^{-4}$   \\ 
 \hline
 (60 , 60) & $3.19\times 10^{-4}$   \\ 
 \hline
 (120 , 120) & $3.56\times 10^{-4}$  \\ 
 \hline
 (180 , 180) & $4.00\times 10^{-4}$  \\ 
 \hline
 (240 , 240) & $2.65\times 10^{-4}$ \\ 
 \hline
 (300 , 300) & $3.14\times 10^{-4}$  \\ 
 \hline
 (360 , 360) & $4.29\times 10^{-4}$  \\ 
 \hline

\end{tabular}
\caption{Relative error with scaled weights.}
\label{accurate_domain_scale_weights}
\end{center}
\end{table}

\begin{table}
\begin{center}
 \begin{tabular}{|| c | c | c  ||} 
 \hline
 ($S_{1,\infty} , S_{2,\infty}$)& $\|\partial L_I/\partial \omega\|_{L^2}$& $\|\partial L_B/\partial \omega\|_{L^2}$   \\ [0.5ex] 
 \hline\hline
 (10 , 10) &$26.372$ & $74.790$   \\ 
 \hline
 (60 , 60) & $949.424$ &$2692.448$   \\ 
 \hline
 (120 , 120) & $3797.696$ & $10769.792$  \\ 
 \hline
 (180 , 180) & $8544.816$  & $24232.031$  \\ 
 \hline
 (240 , 240) & $15190.74$ & $43079.168$ \\ 
 \hline
 (300 , 300) & $23735.602$  & $67311.2$  \\ 
 \hline
 (360 , 360) & $34179.266$ & $96928.125$  \\ 
 \hline

\end{tabular}
\caption{Gradient value for different domains with scaled weights.}
\label{gradient_table_new}
\end{center}
\end{table}


Based on the adapted weights initialization, in Table \ref{fig:Europ_max_call_K}, the results for the European max-call option are presented, and we compare the solution computed by the ANN with the analytical solution given by (\ref{max_call_formula})  for some specific asset prices and different strike values, based on the corresponding loss function and $\lambda=0.5$.  
The parameters considered are, $\sigma_1 = \sigma_2 = 0.2$, $\rho = 0.1$, $\delta_i = 0.1$, $r = 0.05$ and $T = 1$. Moreover, the maximum value of the asset prices is $S_{1\infty} = S_{2\infty} = 4K$. Note that the accuracy of the trained solution is not affected by the size of the domain, in addition, similar to the one-dimensional case, the maximum error is obtained when the underlying value is close to the strike price.

\begin{table}
\begin{center}
 \begin{tabular}{||c | c | c | c ||} 
 \hline
strike & ($S_1$, $S_2$)& ANN & Analytical  \\ [0.5ex] 
 \hline
 &($15,15$) & $-3.16\times 10^{-2} $ & $8.90\times 10^{-5}$ \\ 
15 & ($10,20$) & $4.58\times 10^{-2}$ & $1.16\times 10^{-2}$ \\ 
 &($25,5$) & $2.02\times 10^{-1}$ & $2.11\times 10^{-1}$ \\ 
 \hline
  \hline
& ($30,30$) & $-4.23\times 10^{-2}$ & $1.78\times 10^{-4}$ \\ 
30 & ($20,40$) & $4.75\times 10^{-2}$ & $2.32\times 10^{-2}$ \\ 
 &($50,10$) & $4.10\times 10^{-1}$ & $4.21\times 10^{-1}$ \\ 
 \hline
  \hline
& ($60,60$) & $1.035\times 10^{-1}$ &$3.56\times 10^{-4}$ \\ 
60 & ($40,80$) & $1.79\times 10^{-1}$ & $4.63\times 10^{-2}$ \\ 
 &($100,20$) & $7.79\times 10^{-1}$ & $8.42\times 10^{-1}$ \\ 
 \hline
\end{tabular}
\caption{European max-call option value.}
\label{fig:Europ_max_call_K}
\end{center}
\end{table}

 In Table \ref{error_europ_two_asset}, we present the error for the two-asset European options. The values are computed based on the expression in (\ref{error}).
We observe very fine accuracy for the problems.
\begin{table}
\begin{center}
 \begin{tabular}{||c |c| c ||} 
 \hline
 \text{Option} & $\lambda$ & Error  \\ [0.5ex] 
 \hline\hline
  \text{Asset exchange} & $0.5$ & $4.16\times 10^{-4}$  \\
 \hline
 \text{Max-call $K = 15$} & $0.5$ & $4.55\times 10^{-4}$  \\
 \hline
 \text{Max-call $K = 30$} & $0.5$ & $3.51\times 10^{-4}$  \\
 \hline
 \text{Max-call $K = 60$} & $0.5$ & $3.83\times 10^{-4}$  \\
 \hline
\end{tabular}
\caption{Error according to the loss weight values.}
\label{error_europ_two_asset}
\end{center}
\end{table}

\subsection{American options }
The goal of this section is to address the American option problem by using unsupervised learning with the ANN. As for the European option, we compute the value for one and two underlying assets. However, whereas for the European option, an analytical solution is known, for the American case, we will use the option values computed by finite elements (FEM) using the numerical methods in \cite{Bea} and \cite{Bea-2019} to solve the linear complementarity problem for the American options as the reference.

\begin{figure}[ht!]
  \centering
    \includegraphics[width=0.47\linewidth]{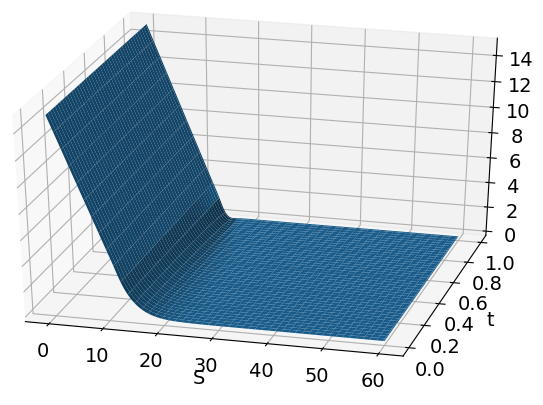}
    \includegraphics[width=0.47\textwidth]
	       {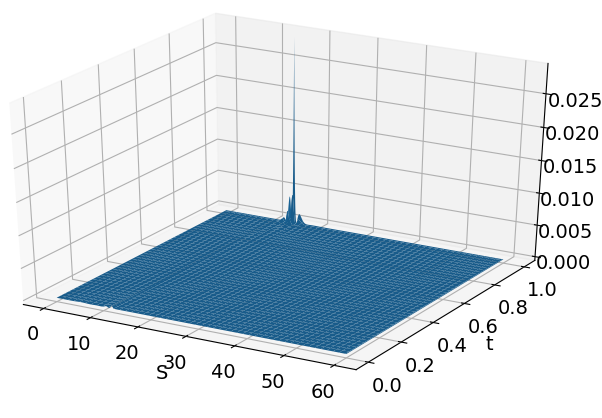}
    \caption{American option price with dividends (left). Error surface comparing with the solution obtained by finite element method (right). Solution obtained by variance normalization method}
  \label{fig:case2_LCP1}
\end{figure}

\begin{figure}[ht!]
  \centering 
    \includegraphics[width=0.45\linewidth]{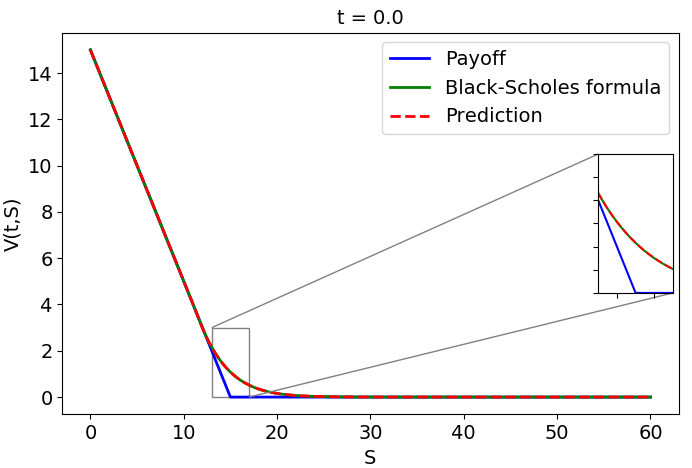}  
    \includegraphics[width=0.45\linewidth]{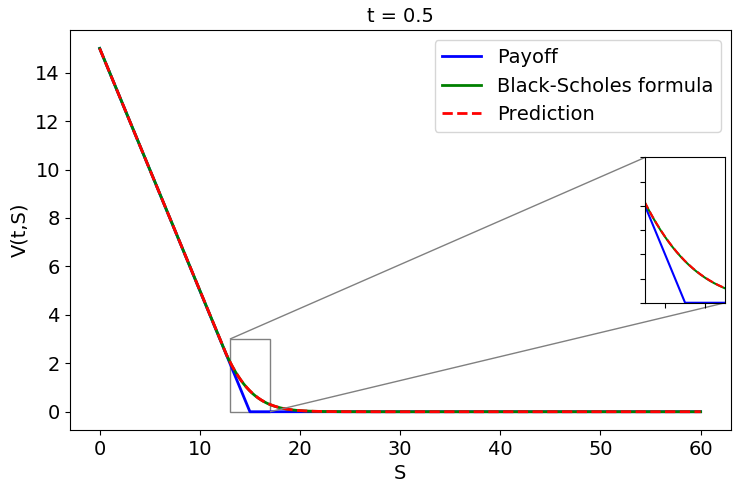}
    \caption{American price and the payoff function. Finite element method (green line) and Neural networks (red line) for different time points. Solution obtained by variance normalization method}
      \label{fig:case2_LCP1_times}
\end{figure}

Similar to the European options problem, the loss function has been optimized using the L-BFGS algorithm, moreover, the ANN is based on the activation function $\tanh(x)$.
With the aim of comparing both methodologies, we price an  American option with the same parameter data that in the previous example,
considering now, the optimal loss weight, which equals $\lambda \approx 0.90$.
We determine first American options with the following parameter data: $\sigma = 0.25$, $\delta = 0.26$, $r = 0.3$, $T = 1$, $K = 15$, $S_\infty = 4K$. Figure \ref{fig:case2_LCP1} shows the trained solution and the error related to a reference FEM solution, for all time points. As for the European options, the maximum error is reached  when the asset price is equal to the strike price, close to the maturity time, where the payoff function is not smooth. In Figure \ref{fig:case2_LCP1_times}, a comparison of the American option value computed by ANNs or FEM is presented for different time points. Moreover, the payoff is plotted to demonstrate that the obstacle condition is satisfied.


\begin{figure}[ht!]
  \centering
    \includegraphics[width=0.47\linewidth]{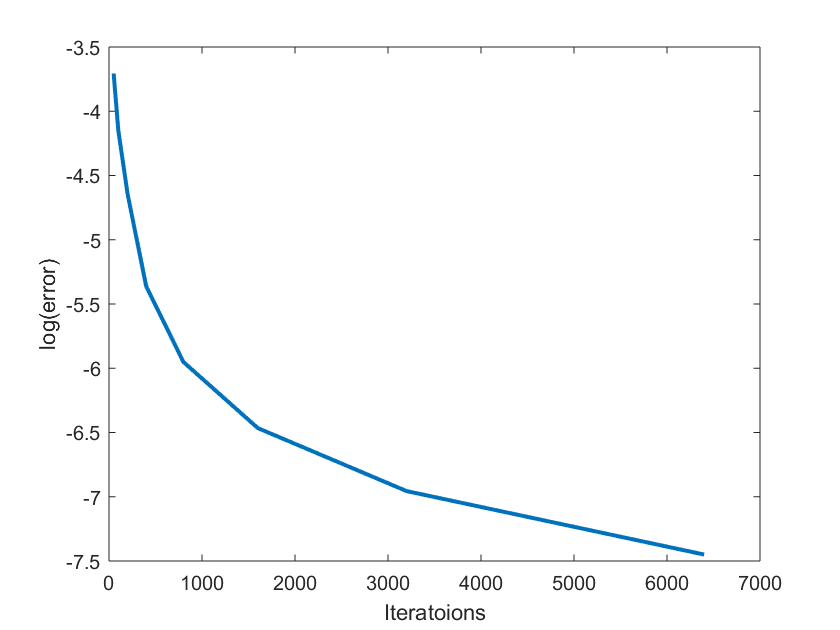}
    \includegraphics[width=0.47\textwidth]
	       {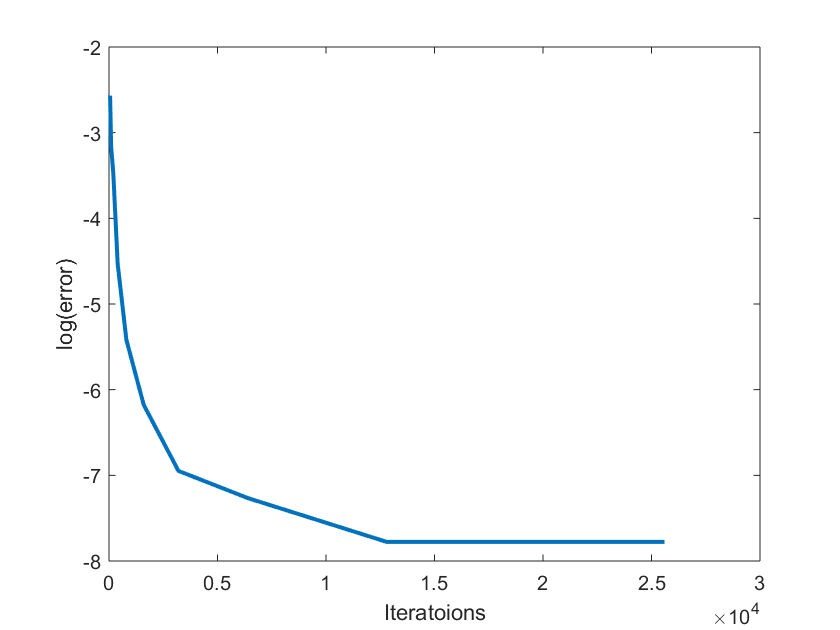}
    \caption{Error value (represented in log scale) obtained for different iterations with the variance normalization method (left) and using the optimal loss weight (right). The reference solution has been obtained solving the PDE by the finite element method.}  
  \label{Convergence}
\end{figure}

The accuracy of two loss functions for the LCP, one based on optimal loss weight and another based on variance normalization,  is compared
by means of the relative error of the solution, computed in terms of the $L^2$-norm, similar to (\ref{error}), i.e.
$$error = \frac{\|v_{FEM}-v_{ANN} \|_{L^2}}{\|v_{FEM}\|_{L^2}}\,.$$
Very similar accuracy is obtained with both loss functions, $5.38\times 10^{-4}$ (for optimal loss weight) versus $5.82\times 10^{-4}$ (variance normalization). However, comparing the convergence of both methodologies, which is presented in Figure \ref{Convergence}, we clearly observe that defining the loss function with a variance normalization (left) the neural network  converges faster than using the optimal loss weight (right). In this figure, the relative error is plotted for different numbers of iterations of the L-BFGS algorithm. 


Next, we value the American options depending on two underlying assets. Optimizing the loss function with the L-BFGS algorithm, with the $\tanh(x)$ as the activation function, and equal weighting of boundary and interior losses, $\lambda=0.5$, the following results have been obtained for the three types of options. 


In Table \ref{table:max_call_Amer}, a comparison between the ANN and FEM solutions is shown. We focus on an American max-call option with several strike values and the following parameter data, $\rho = 0.1$, $\sigma_1 = \sigma_2 = 0.25$, $r = 0.04$, $\delta = 0.01$ and $T = 0.5$. Moreover, for the FEM, $75$ time steps have been considered for the time discretization and the spatial discretization is based on a $101\times 101$ mesh.
\begin{table}
\begin{center}
 \begin{tabular}{||c | c | c | c ||} 
 \hline
strike & ($S_1$, $S_2$)& ANN & FEM  \\ [0.5ex] 
 \hline
 &($15,15$) & $2.021$ & $2.066$ \\ 
15 & ($10,20$) & $5.703$ & $5.643$ \\ 
 &($25,5$) & $10.996$ & $10.969$ \\ 
 \hline
  \hline
& ($30,30$) & $4.102$ & $4.133$ \\ 
30 & ($20,40$) & $11.405$ & $11.29$ \\ 
 &($50,10$) & $21.998$ & $21.938$ \\ 
 \hline
  \hline
& ($60,60$) & $7.916$ &$8.266$ \\ 
60 & ($40,80$) & $22.753$ & $22.573$ \\ 
 &($100,20$) & $43.994$ & $43.877$ \\ 
 \hline
\end{tabular}
\caption{Comparison of American max-call option values. }
\label{table:max_call_Amer}
\end{center}
\end{table}

\begin{figure}[ht!]
  \centering 
    \includegraphics[width=0.45\linewidth]{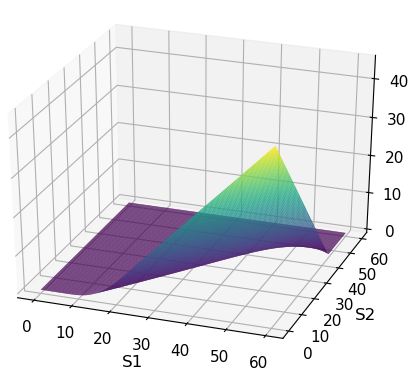}  
    \includegraphics[width=0.45\linewidth]{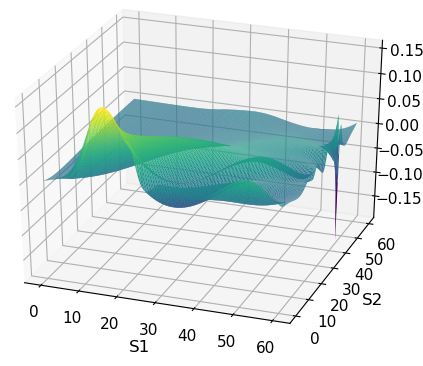}     
    \caption{Two-asset spread American option value in the whole domain (left). Error surface between the FEM and the ANN solution (right).}\label{fig:asset_spread_Amer}
\end{figure}

\begin{figure}[ht!]
  \centering 
    \includegraphics[width=0.45\linewidth]{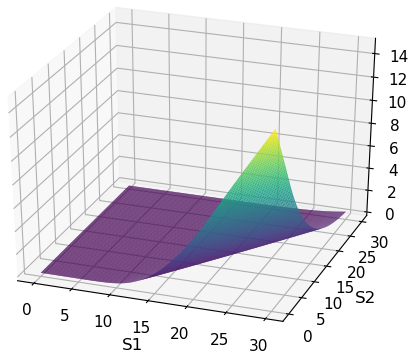}  
    \includegraphics[width=0.45\linewidth]{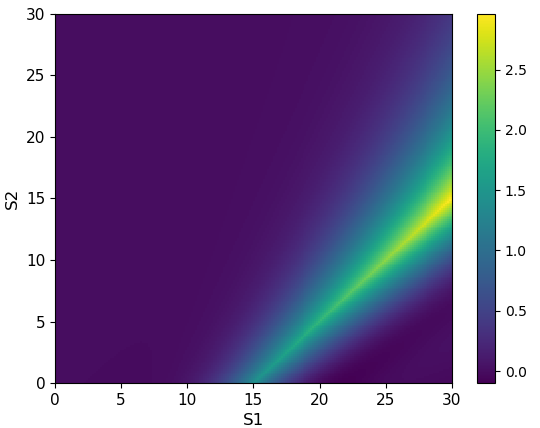}     
    \caption{Two-Asset spread American option value in a reduced domain (left). Difference between the ANN solution and the payoff function (right).}\label{fig:asset_spread_Amer_reduced_domain}
\end{figure}

Figure \ref{fig:asset_spread_Amer} shows the trained solution for the spread American option with the following parameter values,  $K = 15$, $S_{1\infty} = S_{2\infty} = 4K$, $\sigma_1 = \sigma_2 = 0.25$, $\rho = 0.0$, $r_R = 0.04$, $r = 0.3$ and $T= 1$, and the error surface using the FEM solution following \cite{Bea-2019} as a reference. In Figure \ref{fig:asset_spread_Amer_reduced_domain}, the option value and the difference with the payoff function are shown.

%
Finally, in order to show  the accuracy of the method applied to train the ANN to price American options depending on two asset prices, the relative error is presented in Table \ref{Error_table_Amer2d}.
\begin{table}
\begin{center}
 \begin{tabular}{||c | c ||} 
 \hline
  & Error  \\ [0.5ex] 
 \hline\hline
 \text{Max-call} & $1.73\times 10^{-3}$  \\ 
 \hline
 \text{Spread} & $2.45\times 10^{-3}$  \\
 \hline
 \text{Arithmetic average put} & $6.42 \times 10^{-3}$ \\
 \hline
\end{tabular}
\caption{Error for different multi-asset American options}
\label{Error_table_Amer2d}
\end{center}
\end{table}

Note that the accuracy of the neural network for the American options depending on two stochastic factors is lower than for the European options. 
However, this may be because here a numerical solution is our reference and not a closed-form expression. 

\section{Conclusions}

In this work, classical problems in financial option pricing have been addressed with artificial neural networks. In particular, following the classical Black-Scholes model, European and American options depending on one and two underlying assets have been valued. A new unsupervised learning methodology is introduced to solve the option value problems based on the PDE formulation.  With this aim, we  proposed appropriate loss functions. The classical Black-Scholes American option pricing problem has been formulated as a linear complementarity problem. 

For the European option problem, the accuracy of the methods was compared to the analytical solution, whereas, for American options, solutions computed by the finite element method were used as reference values.  For all problems considered, the final error in the ANN solution was highly satisfactory. Needless to mention that ANNs can be easily extended to solving higher-dimensional problems, as they are not drastically affected by the curse of dimensionality. Finally, the PDE problem formulation can be easily generalized by introducing counterparty risk which gives rise to nonlinear option valuation PDEs.

\section*{References}
\bibliographystyle{elsarticle-harv}
\bibliography{xva-references}
\end{document}